\documentclass[useAMS,usenatbib]{mn2e}

\usepackage{graphicx}
\usepackage{natbib}
\usepackage[figuresleft]{rotating}

\newcommand{\ltsima} {$\; \buildrel < \over \sim \;$}
\newcommand{\gtsima} {$\; \buildrel > \over \sim \;$}
\newcommand{\lta} {\lower.5ex\hbox{\ltsima}}
\newcommand{\gta} {\lower.5ex\hbox{\gtsima}}

\newcommand{\Kkmspc}{K km\,s$^{-1}$ pc$^{2}$}
\newcommand{\kms}{km\ s$^{-1}$}

\newcommand{\lya}{Ly$\alpha$}
\title[The Dragonfly Galaxy]{A CO-rich merger shaping a powerful and hyper-luminous infrared radio galaxy at $z=2$: the Dragonfly Galaxy}
\author[B. H. C. Emonts et al.]{B. H. C. Emonts$^{1}$\thanks{Marie Curie Fellow (e-mail:bjornemonts@gmail.com)}, M. Y. Mao$^{2}$, A. Stroe$^{3}$, L. Pentericci$^{4}$, M. Villar-Mart\'{i}n$^{1,5}$, 
\newauthor R. P. Norris$^{6}$, G. Miley$^{3}$, C. De Breuck$^{7}$, G. A. van Moorsel$^{2}$, M. D. Lehnert$^{8}$,  
\newauthor C. L. Carilli$^{2,9}$, H. J. A. R\"{o}ttgering$^{3}$, N. Seymour$^{10}$, E. M. Sadler$^{11}$, R. D. Ekers$^{6}$,   
\newauthor G. Drouart$^{12}$, I. Feain$^{11}$, L. Colina$^{1,5}$, J. Stevens$^{6}$, J. Holt$^{13}$ \\
$^{1}$Centro de Astrobiolog\'{i}a (INTA-CSIC), Ctra de Torrej\'{o}n a Ajalvir, km 4, 28850 Torrej\'{o}n de Ardoz, Madrid, Spain\\
$^{2}$National Radio Astronomy Observatory, P.O. Box 0, Socorro, NM 87801-0387, USA \\
$^{3}$Leiden Observatory, University of Leiden, P.O. Box 9513, 2300 RA Leiden, Netherlands\\
$^{4}$INAF Osservatorio Astronomico di Roma, Via Frascati 33,00040 Monteporzio (RM), Italy\\
$^{5}$Astro-UAM, UAM, Unidad Asociada CSIC\\
$^{6}$CSIRO Astronomy and Space Science, Australia Telescope National Facility, PO Box 76, Epping NSW, 1710, Australia\\
$^{7}$European Southern Observatory, Karl Schwarzschild Strasse 2, 85748 Garching, Germany\\
$^{8}$Institut d'Astrophysique de Paris, UMR 7095, CNRS, Universit\'{e} Pierre et Marie Curie, 98 bis Boulevard Arago, F-75014 Paris\\
$^{9}$Astrophysics Group, Cavendish Laboratory, JJ Thomson Avenue, Cambridge, CB3 0HE, UK\\
$^{10}$International Centre for Radio Astronomy Research, University of Western Australia, 35 Stirling Hwy, Crawley, WA, 6009\\ 
$^{11}$School of Physics, University of Sydney, NSW 2006, Australia\\
$^{12}$Department of Earth and Space Science, Chalmers University of Technology, Onsala Space Observatory, 43992 Onsala, Sweden\\ 
$^{13}$European Space Research and Technology Centre (ESA), Keplerlaan 1, 2201 AZ, Noordwijk, Netherlands\\ 
}
\begin{document}

\date{}

\pubyear{Accepted on 23 April 2015}

\maketitle

\label{firstpage}

\begin{abstract}
In the low-redshift Universe, the most powerful radio sources are often associated with gas-rich galaxy mergers or interactions. We here present evidence for an advanced, gas-rich (`wet') merger associated with a powerful radio galaxy at a redshift of $z$\,$\sim$\,2. This radio galaxy, MRC\,0152-209, is the most infrared-luminous high-redshift radio galaxy known in the southern hemisphere. Using the Australia Telescope Compact Array, we obtained high-resolution CO(1-0) data of cold molecular gas, which we complement with {\it HST/WFPC2} imaging and WHT long-slit spectroscopy. We find that, while roughly M$_{\rm H2} \sim 2 \times 10^{10}$\,M$_{\odot}$ of molecular gas coincides with the central host galaxy, another M$_{\rm H2} \sim 3 \times 10^{10}$\,M$_{\odot}$ is spread across a total extent of $\sim$60 kpc. Most of this widespread CO(1-0) appears to follow prominent tidal features visible in the rest-frame near-UV {\it HST/WFPC2} imaging. \lya\ emission shows an excess over He\,{\small II}, but a deficiency over $L_{\rm IR}$, which is likely the result of photo-ionisation by enhanced but very obscured star formation that was triggered by the merger. In terms of feedback, the radio source is aligned with widespread CO(1-0) emission, which suggests that there is a physical link between the propagating radio jets and the presence of cold molecular gas on scales of the galaxy's halo. Its optical appearance, combined with the transformational stage at which we witness the evolution of MRC\,0152-209, leads us to adopt the name `Dragonfly Galaxy'. 
\end{abstract}


\section{Introduction}
\label{sec:intro}

A crucial epoch in the star formation history of the Universe occurs at $1 < z < 3$, where the evolution of the star formation rate density peaks and the assembly of present-day galaxies is in full progress \citep[e.g.][see \citealt{sha11} for a review]{mad96,lil96,pen01,hop04,hop06,rav06,lot06}. This is also the era in which the quasar density shows a maximum \citep[e.g.][]{sch95,ric06,air10}, indicating that there is a strong link between the evolution of massive galaxies and the growth of their central super-massive black holes. 

While numerical simulations have addressed important aspects of the evolution of massive high-$z$ galaxies, such as hierarchical merging and gas accretion \citep[e.g.][]{spr06,dek09,bou11}, historically, our observational knowledge about galaxy assembly processes at z\,$\ga$\,2 has been dominated by optical and IR studies \citep[e.g.][]{sha11}. While such studies provide important information on the star-formation history of galaxies, they often lack insight into the source that will drive future star formation. For this, we need to investigate the fuel for starburst and AGN activity, namely cold molecular gas.

To date, molecular gas has been detected in a large number of galaxies at $z > 1$, mostly through unresolved detections of the strongest tracer $^{12}$CO({\sl J\,,\,J-1}) (see \citealt{car13}; \citealt{gen15}; and references therein). Significant progress in our understanding of molecular gas at high redshifts has come with the introduction of wide-band receivers at facilities that operate below 50\,GHz, such as the Australia Telescope Compact Array (ATCA) and Karl G. Jansky Very Large Array (VLA). These facilities make it possible to observe the lowest-$J$ transitions of CO, which are among the most robust tracers of the overall molecular gas content, including any low-density and subthermally-excited components \citep[e.g.][]{pap00,hai06,car10,ivi11,rie11extCO,swi11}. By studying the CO-excitation down to the lowest $J$-transitions, indirect results were obtained regarding the presence/absence of widespread molecular gas reservoirs in high-$z$ galaxies \citep[e.g.][]{dad08,dan09,ara10,rie11,dan11,dad14}. Despite this, direct results on the distribution of molecular gas in intermediate- and high-$z$ galaxies are still relatively sparse, although they are growing in number. For example, regularly rotating CO discs have been imaged on scales of up to $\sim 14$\,kpc in various types of high-$z$ galaxies \citep{gen03,gen13,dad10,tac10,tac13,hod12}, while gaseous tidal debris has been mapped in submillimetre galaxies \citep[SMGs;][]{tac08,eng10} and two QSOs \citep{rie08,vil13}.

Among the best studied high-$z$ objects are high-redshift radio galaxies (HzRGs, with $L_{\rm 500\,MHz} > 10^{27}$ W\,Hz$^{-1}$; e.g. \citealt{rot94}). HzRGs are among the most massive and active galaxies in formation, and often reside in galaxy overdensities, as expected from progenitors of current-day giant central-cluster ellipticals \citep[e.g.][]{roc04,ven07,sey07,mil08,bre10,fal10,bar12,wyl13,dro14,hat14,dan14,pod15}. CO studies of HzRGs have revealed systems where molecular gas is spread across 30-40 kpc scales \citep{emo13}, across multiple components \citep[e.g.][]{bre05,ivi08}, in IR-bright companion galaxies \citep{ivi12} or in the galaxy's halo \citep[][]{nes09,emo14}. In addition, a link with the radio-AGN activity was found by \citet{kla04} and \citet[][]{emo14} through alignments seen between the CO and the radio jet axis (Sect.\,\ref{sec:jetsISM}), which resemble jet-alignments seen in optical/UV, X-ray and submm emission \citep{mcc87,cha87,car02,ste03,sma13}. 

Considering these CO results, \citet{ivi12} proposed that gas-rich galaxy mergers are ubiquitous among starbursting HzRGs. 
However, the existing CO studies suffer from a low spatial resolution (often many tens of kpc), hence accurate knowledge about the properties and distribution of molecular gas in HzRGs remains crucially lacking. It is also not clear to what extent other processes, such as filamentary cold-mode accretion of gas from the environment \citep{ker09,dek09,cev10}, may contribute to the gas supply and triggering of activity. Therefore, detailed imaging of the cold molecular gas in HzRGs is essential to further investigate to what extent their starburst and AGN activity is linked to gas-rich galaxy mergers.


In this paper, we present a study of radio galaxy MRC 0152-209 at $z$\,=\,1.92 \citep[P$_{\rm 1.4\,GHz} \sim 1 \times 10^{28}$ W/Hz;][]{pen00radio}. MRC\,0152-209 is the most IR-luminous HzRG known in the southern hemisphere \citep{dro14}. From a detailed analysis of its IR spectral energy distribution, \citet{dro14} found that MRC\,0152-209 has a starburst-IR luminosity of $L_{\rm FIR-SB} \sim 1.8 \times 10^{13} L_{\odot}$, which is in the regime of Hyper-Luminous IR-Galaxies \citep[HyLIRGs, with $L_{\rm FIR} \ge 10^{13} L_{\odot}$; e.g.][]{eis12}. The corresponding star formation rate is estimated to be SFR\,$\sim$\,3000 M$_{\odot}$\,yr$^{-1}$ (Sect.\,\ref{sec:SF}). {\it HST/NICMOS} imaging revealed tail-like features reminiscent of a merger system \citep{pen01}. MRC 0152-209 is thus an ideal target to study the co-evolution of AGN and host galaxy.

In \citet[][hereafter EM11]{emo11a} we published the detection of CO(1-0) emission in MRC\,0152-209 ($L'_{\rm CO} \sim 6.6 \times 10^{10}$ K\,\kms\,pc$^{2}$). We here present new high-resolution ATCA observations that allow us to map the spatial distribution of the CO(1-0). We will show that MRC\,0152-209 is one of very few high-$z$ galaxies with observational evidence for tidal debris of cold molecular gas on a scale of $\sim$60 kpc. Notably, the radio source is aligned with widespread CO(1-0) emission, suggesting a link between the cold molecular gas and the propagating radio jets. We also present new {\it HST/WFPC2} imaging and WHT long-slit spectroscopy to further investigate the violent merger episode that we witness in MRC\,0152-209.

Throughout this paper, we assume H$_{0}$=71 \kms\ Mpc$^{-1}$, $\Omega_{\rm M}$=0.3 and $\Omega_{\Lambda}$=0.7, corresponding to an angular scale of 8.3 kpc arcsec$^{-1}$ for MRC~0152-209 \citep[][]{wri06}.\footnote{See http://www.astro.ucla.edu/$\sim$wright/CosmoCalc.html}

\section{Observations}
\label{sec:observations}

\subsection{CO(1-0)}
\label{sec:obsCO}

Observations of CO(1-0) were performed with the ATCA. We combined existing data taken on 25,\,26 Aug. $\&$ 28,\,29 Sept 2010 with the most compact hybrid H75 and H168 array configurations (baselines ranging from $31-192$ meters; EM11) with new observations done on 23\,-\,25 June $\&$ 14\,-\,16 May 2011 using the inner 5 antennas of the 750B and 1.5B array configurations (covering baselines up to 1.3 km). Because phase stability decreases when baseline length increases or weather conditions deteriorate, we only used data taken under good weather conditions, i.e., data for which the root-mean-square path-length of the atmospheric seeing was an average [maximum] of 80 [200] $\mu$m for the 750m and 1.5k array configurations and an average [maximum] of 170 [300] $\mu$m for the compact hybrid configurations \citep[see][]{mid06}.\footnote{Quoting \citet{mid06}: ``The ATCA seeing monitor is a [2-dish] interferometer on a 230m east-west baseline that tracks the 30.48\,GHz beacon of a geostationary communications satellite at elevation 60$^{\circ}$''} The total on-source integration time under these conditions was 40h (14.9h in 1.5k array, 12.6h in 750m array, 5.1h in H168 array and 7.8h in H75 array).

The observational setup, calibration and data reduction strategy was the same as described in \citet{emo11a,emo11b}. We used the Compact Array Broadband Backend \citep{wil11}, with two 2\,GHz bands and 1\,MHz channels. We centred both bands around $\nu = 39.46$\,GHz, which is the frequency of the redshifted CO(1-0) line ($\nu_{\rm rest} = 115.2712$\,GHz). We took 2\,min scans on PKS\,B0130-171 to calibrate the phases and bandpass every 5\,-\,10 min, with the interval depending on the weather conditions and array configuration. Uranus was used to calibrate the absolute flux. Given that atmospheric phase instabilities increase significantly when going below 30$^{\circ}$ in elevation, most of our observations were scheduled to avoid very low elevations. Because the ATCA only has east-west baselines in its extended 750m and 1.5k array configurations, this resulted in a lack of long-baseline uv-coverage at large hour angles, and thus an elongated beam with significantly worse spatial resolution in NS compared to EW direction. 

The data were calibrated in MIRIAD \citep{sau95} following EM11. We used a robustness parameter of 0.5 \citep{bri95} to image the data that are presented in this paper. This resulted in a beam-size of $4.0'' \times 1.3''$ (PA = -4.4$^{\circ}$). The uncertainty in absolute flux calibration is $\sim$30$\%$ (EM11). The final data set (covering $\sim 15,000$\,\kms\ with a full resolution of 7.6 \kms\ per 1\,MHz channel) was binned to a channel width of 30\,\kms\ and subsequently Hanning smoothed to an effective velocity resolution of 61 \kms, with a noise level of $\sigma=0.146$ mJy\,beam$^{-1}$\,chan$^{-1}$. The data in this paper are presented in optical barycentric velocity relative to the redshift of 
$z = 1.9212$ that we derived from the total CO(1-0) emission-line profile in EM11.

\subsection{HST imaging}
\label{sec:datahst}

An optical image of MRC\,0152-209 was taken with the {\it Hubble Space Telescope (HST) Wide Field and Planetary Camera 2 (WFPC2)} on 31 Oct 2000 (Project $\#$8183, PI: G.\,Miley). The object was placed in the Planetary Camera frame, which has a pixel size of 0.0455 arcsec. The F555W broad-band filter ($4790 - 6020$\,\AA) was used, which covers the near-UV rest-frame ($1640 - 2062$\,\AA) for $z=1.92$. The F555W filter also includes the redshifted He\,II$_{1640}$ and C\,III]$_{\rm 1909}$. However, from the spectrum that will be discussed in Sect.\,\ref{sec:resultswht}, we estimate that the contribution of He\,II and C\,III] to the total flux in the broad-band filter is $<$10$\%$. 

We aligned the {\it HST/WFPC2} image with the radio data by assuming that the central host galaxy is co-spatial with the peak of the CO(1-0) emission. This CO(1-0) peak-emission also coincides with the brightest component in the radio continuum from \citealt{pen00radio} (Sect.\,\ref{sec:resultsco}). The resulting {\it HST} astrometry agrees with the 3.6$\mu$m peak in existing Spitzer IRAC channel 1 data from \citealt{sey07} (which has an estimated uncertainty in absolute astrometry of $\sim$0.3$^{\prime\prime}$). However, because the exact radio core is not known from existing radio continuum data, we conservatively assume that the astrometric uncertainty in our {\it HST} overlay is $\sim$0.5 arcsec. 

We also updated the astrometry of the near-IR {\it HST/NICMOS} F160W image from \citet{pen01} by matching its nuclear features to the same features seen in the {\it WFPC2} image. This NICMOS image covers the $4840 - 6230$\,\AA\ restframe emission at $z=1.9212$.

\subsection{Spectroscopy}
\label{sec:WHT}

A long-slit spectrum of MRC\,0152-209 was taken on 25 July 2012 using the Intermediate-dispersion Spectrograph and Imaging System (ISIS) on the William Herschel Telescope (WHT), La Palma. We used the R300B/R316R grating on the blue/red arm, the D5300 dichroic and the GG495 blocking filter in the red arm to cut out second order blue light. This setup resulted in an effective wavelength coverage of $3450-7270$\,\AA\ ($1181-2490$\,\AA\ in the rest-frame), with a $\lambda$-resolution of $\sim$5\AA. The slit had a width of 1.3 arcsec. Given that the final CO(1-0) results were not available at the time of the WHT observations, the slit was positioned mid-way between the two 8.2\,GHz radio continuum components imaged by \citet{pen00radio}, at RA=01:54:55.74 and dec=-20:40:26.42. Because of the low elevation at which the observations had to be done (airmass 2.4), we positioned the slit at the parallactic angle of PA\,=\,320$^{\circ}$. The on-source integration time was 3600\,sec, divided into $6\times600$\,sec consecutive exposures. 

For the data reduction we used the Image Reduction and Analysis Facility \citep[IRAF;][]{tod93}. A standard calibration was done on each frame, including bias subtraction, flatfielding and wavelength calibration. For the latter, we used an arc exposure taken at approximately the same time and elevation as the target spectra. The frames were then combined to remove cosmic rays and the background was subtracted. Four standard stars (Feige\,98, Feige\,110, BD+253941 and Hiltner\,102) were used for flux calibration.

For the analysis of the spectra, we used the Starlink packages FIGARO and DIPSO. We binned consecutive rows of pixels to create a 1-D spectrum that contains the integrated flux across the full region were line-emission from MRC\,0152-209 was detected. This spectrum was analysed by fitting Gaussian profiles to the emission lines. The spectrum is shown in the observed wavelength frame in this paper.

\section{Results}
\label{sec:results}

\subsection{Stellar/HST morphology}
\label{sec:resultshst}

\begin{figure*}
\centering
\includegraphics[width=0.7\textwidth]{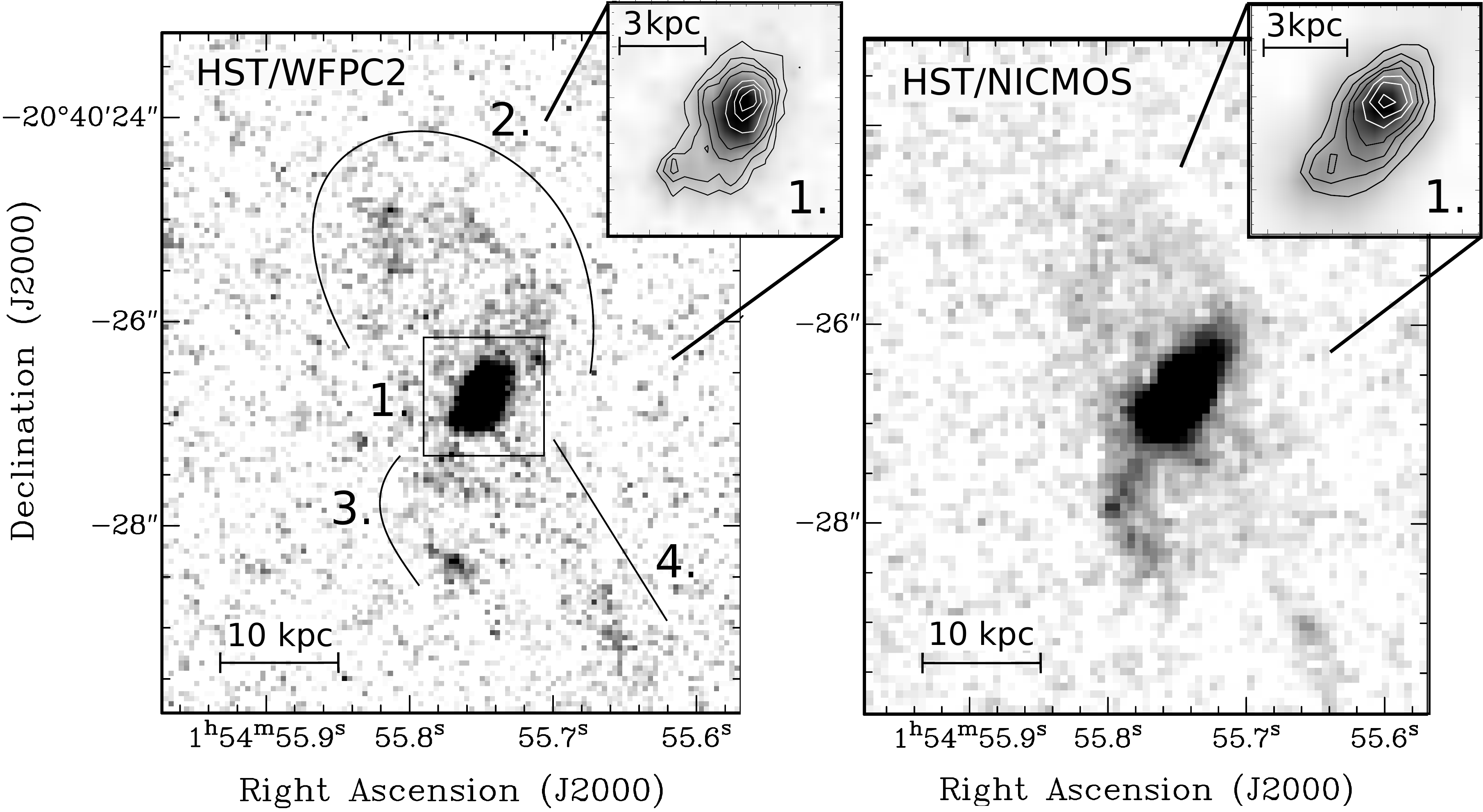}
\caption{{\sl Left:} {\it HST/WFPC2} F555W image of MRC\,0152-209. The four distinct features are described in the text. The inset on the top show a zoom-in of the central region $\#$1. {\sl Right:} {\it HST/NICMOS} F160W image at the same scale as the {\it HST/WFPC2} image. The NICMOS data have been reproduced from Pentericci et al (2001), their Fig.\, 2. $\texttt{\copyright}$ AAS. Reproduced with permission.}
\label{fig:hst}
\end{figure*}

Figure \ref{fig:hst} shows the {\it HST} imaging of MRC\,0152-209. We argue that MRC\,0152-209 is a system that is undergoing a major merger, based on several distinct features: {\sl 1).} A double nuclear component (separation 0.44$^{\prime\prime}$ or 3.7\,kpc). This double component is visible both in the {\it WFPC2} and {\it NICMOS} imaging and could indicate the presence of either a close double nucleus or central dust feature; {\sl 2).} A broad plume-like feature stretching 20\,kpc (2.4$^{\prime\prime}$) north of the center. From the {\it WFPC2} image, this feature appears to be a tidal arm looping back onto the host galaxy; {\sl 3).} A southern tidal arm/tail that is most prominent in the {\it NICMOS} image. In the middle of the tidal arm (at roughly 10\,kpc distance from the center) there is an apparent {\it NICMOS} companion that is not visible in the {\it WFPC2} image. Both this apparent companion and the middle part of the tidal arm therefore appear to be heavily obscured in the near-UV restframe; {\sl 4).} A faint SW tail that ends in a region with enhanced emission in both {\it NICMOS} and {\it WFPC2}, roughly 23\,kpc (2.8$^{\prime\prime}$) from the center. All these features are indicated in Fig.\,\ref{fig:hst}.

\begin{figure*}
\centering
\includegraphics[width=0.98\textwidth]{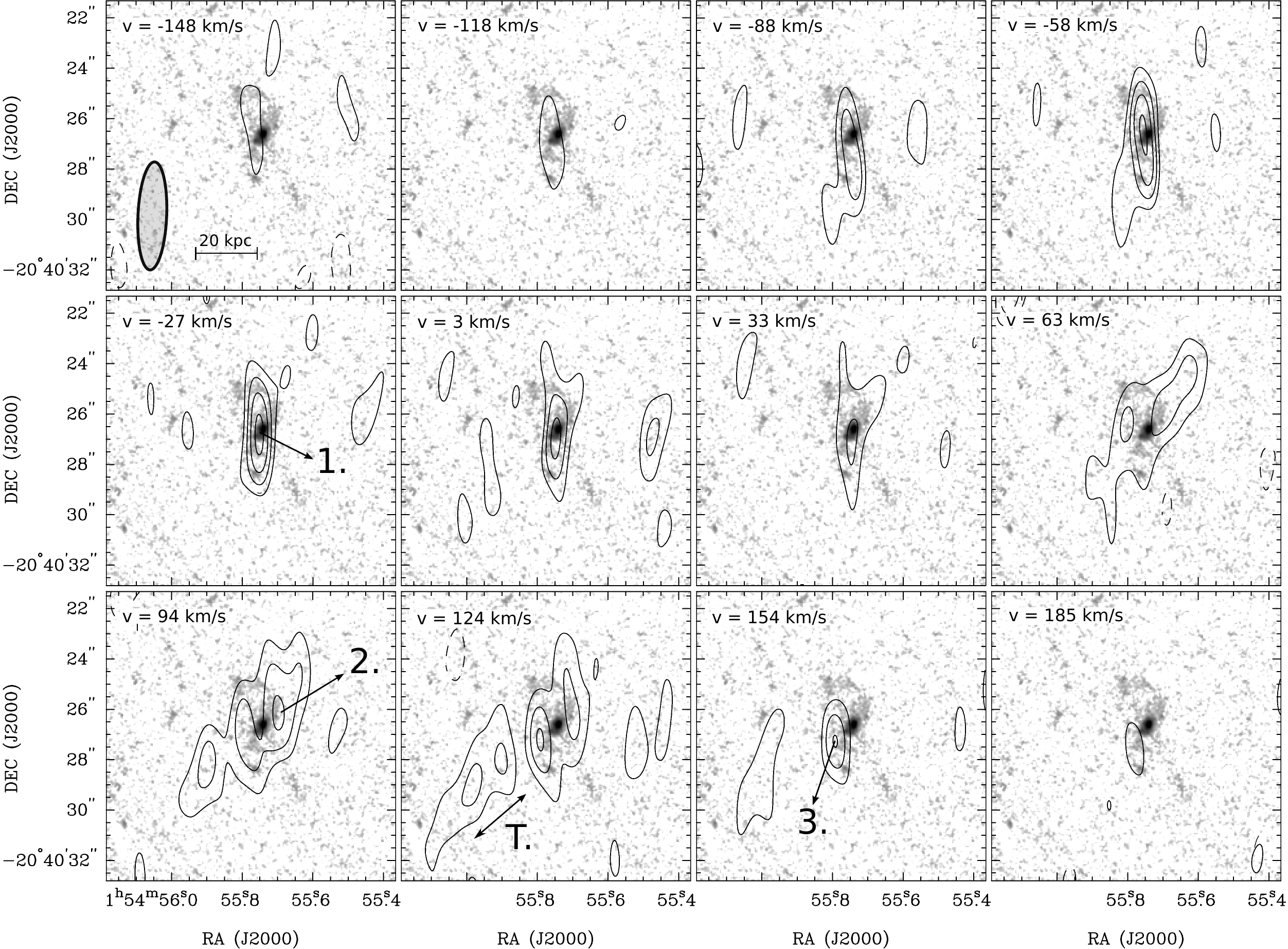}
\caption{Channel maps of the CO(1-0) emission in MRC\,0152-209. Contours of CO(1-0) emission are overlaid onto the {\it HST/WFPC2} image of Fig.\,\ref{fig:hst}. Contour levels: -3, -2 (dashed), 2, 3, 4, 5 (solid) $\times \sigma$, with $\sigma = 0.146$ mJy\,beam$^{-1}$ the root-mean-square (rms) noise level. The distinct regions indentified in the plot correspond to the same regions as in the {\it HST} image of Fig.\,\ref{fig:hst}, with ``T'' an additional tail-like feature that is only seen in CO(1-0). Note the highly elongated beam (ellipse) that we obtained by using the extended East-West array configurations of the ATCA (see Sect.\,\ref{sec:obsCO} for details).}
\label{fig:COdistribution}
\end{figure*}

\begin{figure*}
\centering
\includegraphics[width=0.98\textwidth]{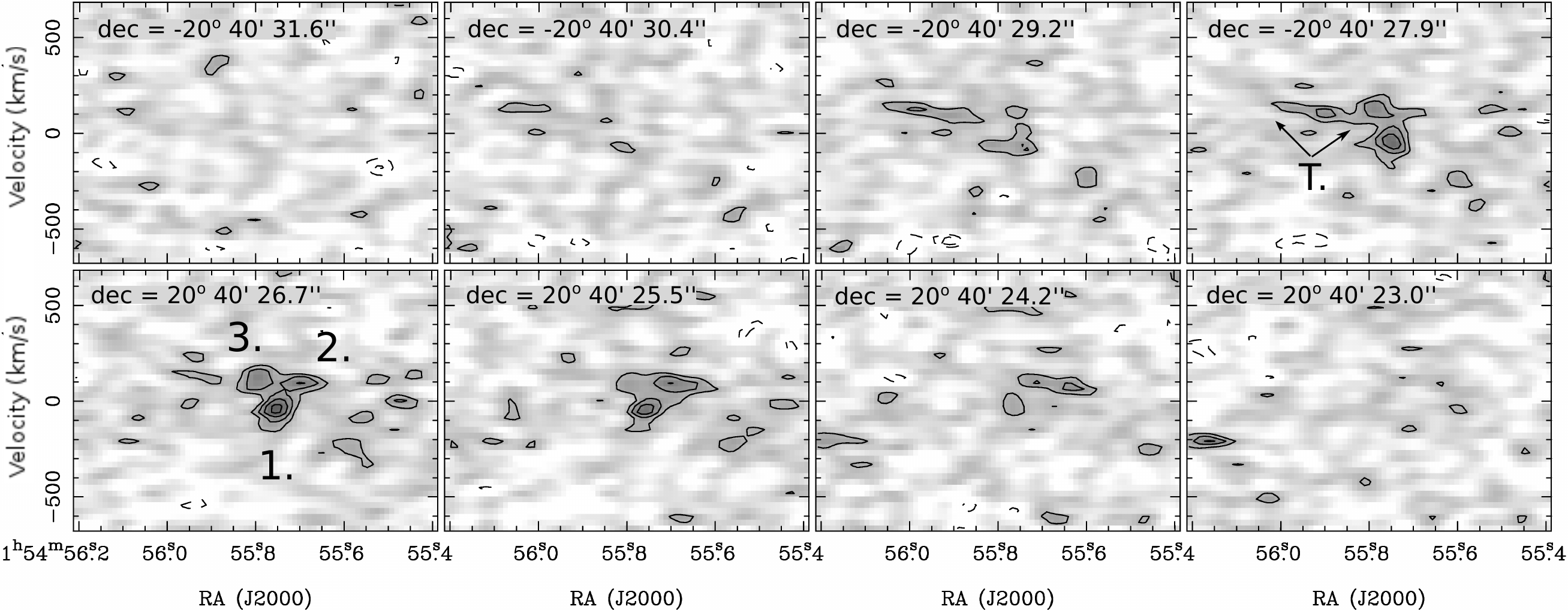}
\caption{Channel maps of the CO(1-0) emission, shown in RA vs. velocity. The data are binned to 1.2$^{\prime\prime}$ in declination. Contour levels: -3, -2 (dashed), 2, 3, 4, 5 (solid) $\times \sigma$, with $\sigma = 0.143$ mJy\,beam$^{-1}$ the rms noise level. The distinct regions indentified in the plot correspond to the same regions as in Fig.\,\ref{fig:COdistribution}.}
\label{fig:COdistributionXZ}
\end{figure*}

\begin{figure*}
\centering
\includegraphics[width=\textwidth]{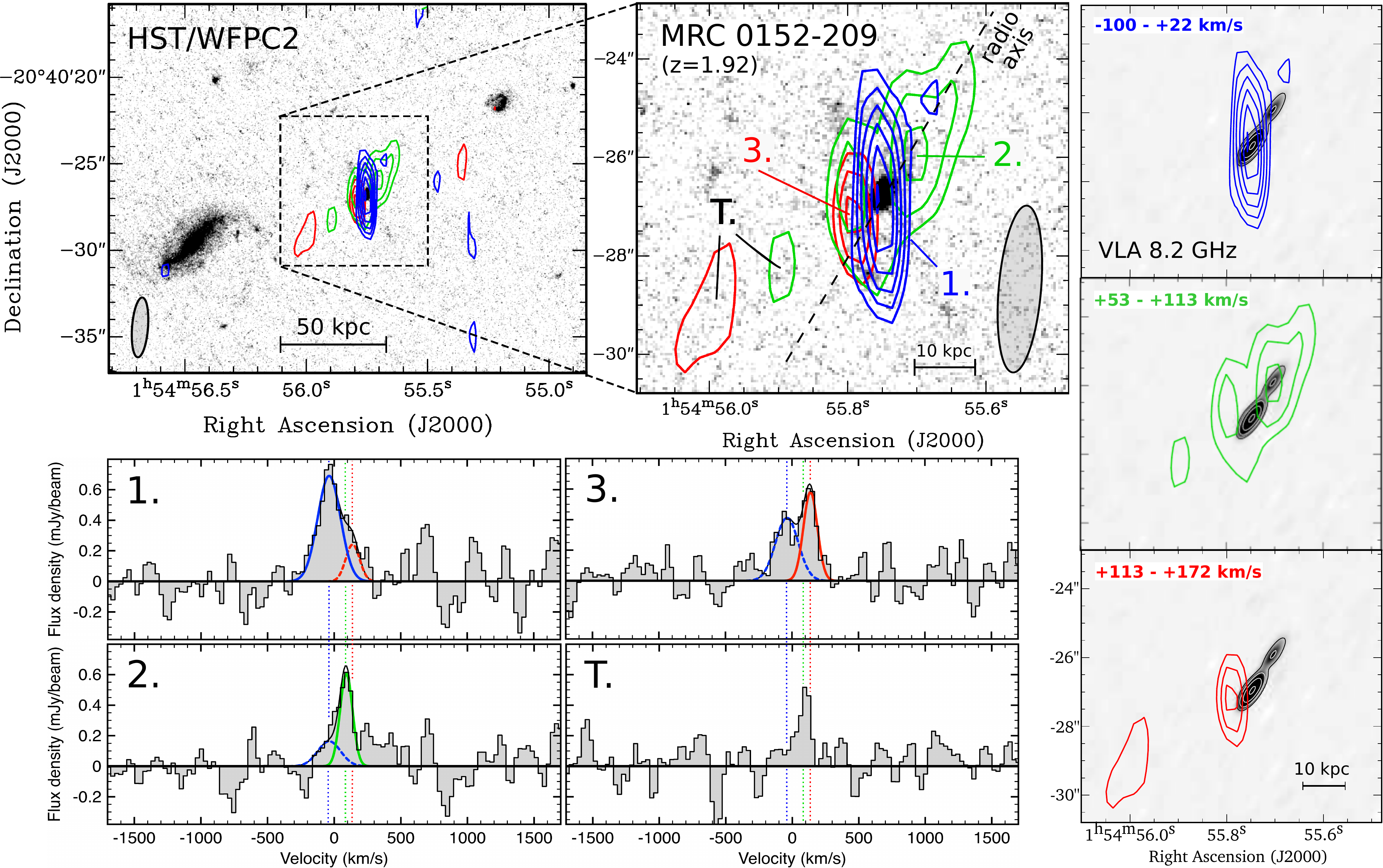}
\caption{{\sl Top left:} Total intensity image of MRC\,0152-209 integrated over three distinct velocity ranges (blue = -100\,$-$\,+22 \kms; green = 53\,$-$\,113 \kms; red = 113\,$-$\,172 \kms) and overlaid onto the {\it HST/WFPC2} image of Fig.\,\ref{fig:hst} to within an accuracy of $\la$0.5$^{\prime\prime}$ (part of this figure is reproduced, with permission, from \citealt{emo14}, their Fig.\,4). CO(1-0) contour levels are at 2.8, 3.5, 4.2, 4.9, 5.6, 6.3$\sigma$ level (with $\sigma$ = 12 mJy\,beam$^{-1}$\,$\times$\,\kms\ for the blue component and 7.3 mJy\,beam$^{-1}$\,$\times$\,\kms\ for the red/green components). The dashed line visualises the extrapolated radio axis. {\sl Right:} Plots of the individual total intensity maps, as extracted across the three different velocity ranges. The 8.2\,GHz radio continuum source from \citet{pen00radio} is overlaid in greyscale and thin contours (levels at 2, 7, 25 and 86$\%$ of the peak intensity). {\sl Bottom:} CO(1-0) emission-line spectra in the four distinctive regions indicated in the total intensity plot on the top. The data are Hanning smoothed and the velocity is shown with respect to $z=1.9212$ (Sect.\,\ref{sec:observations}). The solid blue, green and red lines represent Gaussian fits to the CO emission from regions 1, 2 and 3, respectively. From the flux covered by these Gaussians we derived $L'_{\rm CO}$ for each region. The dashed colored lines indicate the estimated spill-over emission from adjacent regions that are separated by one beam-size or less (this emission is not included in the determination of  $L'_{\rm CO}$). The thin black line is the overall fit. {\sl [A color version of this Figure is available in the online edition of the paper.]}}
\label{fig:mrc0152color}
\end{figure*}

\subsection{Molecular CO(1-0) gas}
\label{sec:resultsco}

In EM11 we presented the detection of CO(1-0) emission in MRC\,0152-209, with $L'_{\rm CO(1-0)} \sim 6.6 \times 10^{10} L_{\odot}$. Figures \ref{fig:COdistribution} and \ref{fig:COdistributionXZ} show the distribution of CO(1-0) in MRC\,0152-209 that we derive from our new high-resolution ATCA observations. Part of the CO(1-0) emission appears to be associated with the bright optical emission of the central host galaxy (with `host galaxy' defined in this paper as the region of bright optical emission that covers the double nuclear component in Fig.\,\ref{fig:hst}; i.e. region $\#$1). However, a large fraction of the CO(1-0) emission is clearly off-set from the host galaxy. This gas appears to coincide with several of the tidal features seen in the {\it HST/WFPC2} and {\it HST/NICMOS} imaging, namely the inner parts of the looped northern arm (region $\#$2) and the southern arm/tail (region $\#$3). Following \citealt[][]{pap08} (see also \citealt{fol99}), we estimate that the astrometry of the CO emission-line peak in regions 2 and 3 has an accuracy of $\delta \theta_{\rm rms} \sim \frac{1}{2} \langle \Theta_{\rm beam} \rangle {\rm (S/N)}^{-1} \sim 0.5^{\prime\prime} {\rm (NS)} \times 0.2^{\prime\prime} {\rm (EW)}$, for $\Theta_{\rm beam} \sim 4^{\prime\prime} \times 1.3^{\prime\prime}$ and S/N $\sim$ 4.5. This confirms that part of the CO(1-0) is indeed off-set by at least 20\,kpc from the central host galaxy. There is also a weak tail of redshifted CO(1-0) that is not detected in the {\it HST} image and that stretches up to $\sim$35 kpc (4$^{\prime\prime}$) SE of the host galaxy (region ``T'' in Figs.\,\ref{fig:COdistribution}$-$\ref{fig:mrc0152color}). We argue that the CO(1-0) in regions 2, 3 and ``T'' is most likely tidal-debris of cold molecular gas that follows the tidal features visible in the {\it HST} imaging, and which is spread across a total extent of $\sim$60 kpc. Alternatively, some of the extended CO(1-0) components may identify gas-rich merging companion galaxies. We do not favor this scenario, because no identifiable candidate galaxies are found at the location of the extended CO(1-0) in the {\it HST/WFPC2} and {\it NICMOS} imaging. However, given the $\sim$0.5$^{\prime\prime}$ uncertainty in absolute astrometry between the ATCA and {\it HST} data, and the presence of a possible {\it HST/NICMOS} companion $\sim$10\,kpc towards the South-East (Fig.\,\ref{fig:hst}), we cannot completely rule out this possibility with our current data.

Figure\,\ref{fig:mrc0152color} shows the total intensity maps and spectra of the various CO(1-0) components. We derive CO luminosities of the three main components (1, 2 and 3) from their emission-line spectra. We use this approach because the CO emission from these three components is not completely mutually independent (given that their spatial separation is on the order of one beam-size), and in the spectra any contribution from adjacent components can be more easily identified than in the total intensity plots (Fig.\,\ref{fig:mrc0152color} - bottom). We estimate that $L'_{\rm CO} = 2.7\,\pm\,0.3 \times 10^{10}$ \Kkmspc\ is associated with the central host galaxy (the ``blue'' component $\#$1 in Fig\,\ref{fig:mrc0152color}), while $L'_{\rm CO} = 1.2\,\pm\,0.3 $ and $1.4\,\pm\,0.3 \times 10^{10}$ \Kkmspc\ are associated with the suggested tidal debris of the green ($\#$2) and red ($\#$3) component, respectively.\footnote{Uncertainties quoted in this Section were calculated based on the FWZI of the Gaussian fits in Fig.\,\ref{fig:mrc0152color} (bottom), following Equation 2 from \citet[][based on \citealt{sag90}]{emo14}. They do not reflect the additional 30$\%$ error in absolute flux calibration (Sect.\,\ref{sec:obsCO}), because this is a scaling factor that affects all components in the same way.} Combined, these three components recover roughly 80$\%$ of the total CO(1-0) luminosity from the low-resolution observations of EM11 ($L'_{\rm CO} \sim 6.6 \pm 0.8 \times 10^{10}$ \Kkmspc; see also \citealt{emo14}). This suggests that another 20$\%$ of the CO emission ($L'_{\rm CO} \sim 1.3\,\pm\,0.9 \times 10^{10}$ \Kkmspc) is associated with even more widespread gas, such as the green extension NW of component 2 and the tentative tail towards the SE. Therefore, $\sim$40$\%$ of the overall CO(1-0) emission in MRC\,0152-209 (i.e. $L'_{\rm } = 2.7 \pm 0.3 \times 10^{10}$ \Kkmspc) appears to be associated with the central host galaxy, while $\sim$60$\%$ ($L'_{\rm } = 3.9 \pm 1.0 \times 10^{10}$ \Kkmspc) is found across the widespread tidal debris. Fig.\,\ref{fig:mrc0152color} also shows that the tidal CO(1-0) is found at higher velocities than the CO(1-0) in the central host galaxy. 

The 8.2\,GHz radio continuum source appears to be aligned with the CO-emitting gas (Fig.\,\ref{fig:mrc0152color} - right). In particular the alignment with the widespread CO(1-0) emission NW of the host galaxy is striking. \citet{pen00radio} classify the radio source as a small double-lobed radio source with two steep-spectrum hot-spots ($\alpha^{8.2}_{4.7} = -1.4$ for the NW and -1.7 for the SE hot-spot), which cover a total linear extent of $\sim$18 kpc (2.2$^{\prime\prime}$). We are currently in the process of imaging the structure of the radio source at much higher resolution with new observations. If confirmed that this is a jet, the alignment with the widespread CO(1-0) emission NW of the nucleus would suggests that there is a direct physical link between the cold molecular gas and the propagating radio source. We will discuss this further in Sect.\,\ref{sec:jetsISM}.

\subsection{Near-UV emission-line gas}
\label{sec:resultswht}

\begin{figure*}
\centering
\includegraphics[width=\textwidth]{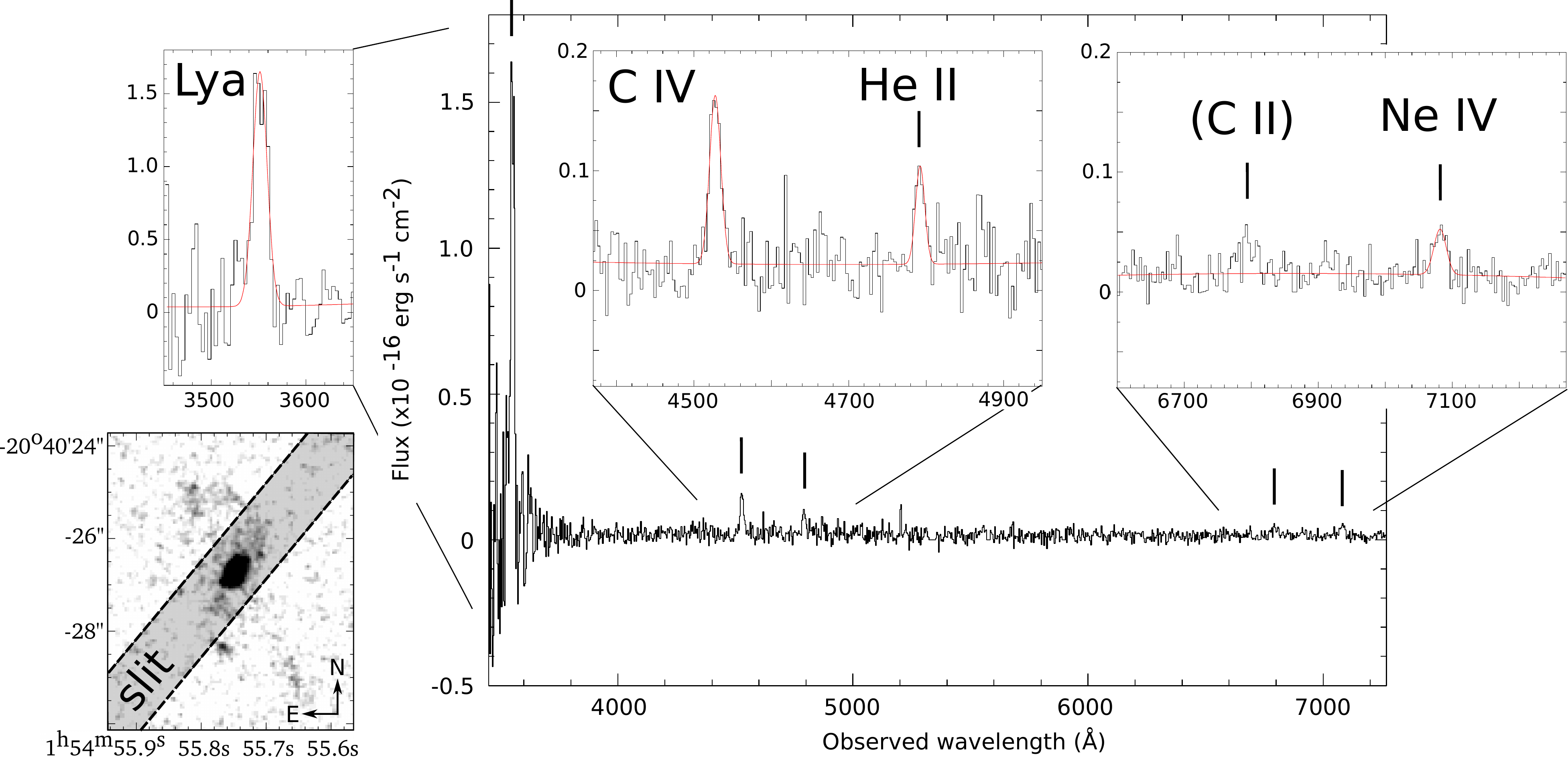}
\caption{WHT long-slit spectrum of MRC\,0152-209. The bottom-left plot visualizes the slit alignment (see Sect.\,\ref{sec:WHT} for details). The lines of \lya, C\,{\small IV}, He\,{\small II} and Ne\,{\small IV} are clearly detected, with a tentative detection of C\,{\small II}. The red line in the insets shows a Gaussian fit to the line-profiles.}
\label{fig:spectrum}
\end{figure*}

\begin{table*}
\caption{Optical spectroscopy. Observed redshift $z$, FWHM (corrected for instrumental broadening), line flux/luminosity and relative strength with respect to \lya\ for the various emission lines detected in the optical spectrum. For completion, the $z$ and FWHM of the CO(1-0) line emission form EM11, based on the integrated emission-line profile, are also included.}
\label{tab:spec}
\begin{tabular}{lccccc}
\hline
\hline
Line & $z$  & FWHM (\kms) & S$_{\rm line}$ ($10^{-16}$\,erg\,s$^{-1}$\,cm$^{-2}$) & $L_{\rm line}$ ($10^{42}$\,erg\,s$^{-1}$) & R\,(S$_{\rm Ly\alpha}$/S$_{\rm line}$) \\ 
& & & & & \\
Ly$\alpha$$_{1215.7{\rm \AA}}$ & 1.9215\,$\pm$\,0.0007 & 1301\,$\pm$\,187 & 30.5\,$\pm$\,2.8 & 81.8\,$\pm$\,7.5 & 1.0 \\
C\,{\small IV}$_{1548+1551{\rm \AA}}$ & 1.922\,$\pm$\,0.02 & 900\,$\pm$\,200 & 2.56\,$\pm$\,0.24 & 6.87\,$\pm$\,0.64 & 11.9 \\
He\,{\small II$_{1640.4{\rm \AA}}$} & 1.9214\,$\pm$\,0.0007 & 687\,$\pm$\,147 & 1.23\,$\pm$\,0.20 & 3.30\,$\pm$\,0.54 & 24.8 \\
Ne\,{\small IV}$_{2424.4{\rm \AA}}$ & 1.9211\,$\pm$\,0.0007 & 839\,$\pm$\,148 & 0.88\,$\pm$\,0.14 & 2.36\,$\pm$\,0.38 & 34.7 \\
CO(1-0)$_{115.3\,{\rm GHz}}$ & 1.9212\,$\pm$\,0.0002 & 397\,$\pm$\,37 & - & - & - \\
\hline
\end{tabular} 
\flushleft 
\end{table*} 

Figure \ref{fig:spectrum} shows the WHT optical long-slit spectrum of MRC\,0152-209, which covers $1181-2490$\,\AA\ in the near-UV rest-frame. We clearly detect the emission lines of \lya, C\,{\small IV}, He\,{\small II} and Ne\,{\small IV}, with a tentative detection of C\,{\small II}. Our measured \lya\ flux is in good agreement with the value quoted by \citet{mcc91}. The CIII] line, as reported by \citet{mcc91}, fell on a prominent broad sky-line residual in our data (which we blanked-out in Fig.\,\ref{fig:spectrum}), hence could not be detected. The properties of these emission lines are summarised in Table \ref{tab:spec}.

The emission lines are all well approximated by a single Gaussian profile (Fig.\,\ref{fig:spectrum}). The FWHM of the emission lines (corrected for instrumental broadening) varies from 1300 \kms\ for the strong \lya\ line to $\sim$700 \kms\ for He\,{\small II}, compared to FWHM\,$\sim$\,400 \kms\ for the integrated CO(1-0) profile (EM11).\footnote{The CO(1-0) profile from EM11 was extracted across a larger region than the WHT spectrum. Fig.\,\ref{fig:spectrum} shows that the near-UV emission from the central host galaxy was included in the slit, but that some of the extended emission may have been missed. Therefore, the FWHM of the near-UV lines are firm lower limits.} The relatively broad \lya\ profile can partially be explained by resonant scattering. Nevertheless, even for the other emission lines of ionised gas, the line-widths are roughly twice that of the CO(1-0) emission. 
This is similar to what has been found when comparing the [O\,{\small III}] and CO kinematics in low-$z$ QSOs, and has been ascribed to a different geometry and spatial extent for the ionised and molecular gas phase, as well as a higher sensitivity of the ionised gas to non-gravitational motions, such as outflows \citep{shi06,vil13sample}. No clear absorption features are observed in any of the emission lines. We note, however, that the \lya\ line falls at the edge of our spectrum, where a significant drop in sensitivity resulted in an increased noise level.

To within the uncertainty, the redshift derived from each of the optical emission lines is in excellent agreement with $z = 1.9212 \pm 0.0002$ derived from the integrated CO(1-0) profile in EM11. 

\section{Discussion: the `Dragonfly Galaxy'}
\label{sec:discussion}

We presented high-resolution CO(1-0) data, complemented with {\it HST/WFPC2} imaging and WHT spectroscopy, of the merger system MRC\,152-209 ($z$\,=\,1.92). Based on the optical morphology of the galaxy in our {\it HST/WFPC2} image (Fig.\,\ref{fig:hst}), we will here adopt the name `Dragonfly Galaxy'. In this Section we will further discuss the CO(1-0) properties (Sect.\,\ref{sec:molecular}), compare those to the extreme star formation rates (Sect.\,\ref{sec:SF}) and investigate AGN feedback (Sect.\,\ref{sec:jetsISM}). This will further reveal the evolutionary stage of this object. 


\subsection{Molecular gas $\&$ fueling}
\label{sec:molecular}

We saw in Sect.\,\ref{sec:resultsco} that a large amount of cold molecular gas ($L'_{\rm CO} \sim 3.9 \pm 1.0 \times 10^{10}$ K\,\kms\,pc$^{2}$) appears to reside in large-scale tidal features that stretch several tens of kpc from the center of the Dragonfly Galaxy. We assume a standard conversion factor of M$_{\rm H2}$/$L'_{\rm CO} = 0.8$ M$_{\odot}$ (\Kkmspc)$^{-1}$ to translate our CO(1-0) estimates into molecular gas masses. This agrees with M$_{\rm H2}$/$L'_{\rm CO}$ found in ultra-luminous infra-red galaxies \citep[][]{dow98}, as well as high-$z$ sub-millimeter and starforming galaxies \citep{tac08,sta08}, although we discussed in EM11 that this conversion has a large uncertainty. This means that M$_{\rm H2} \sim 3.1 \pm 0.8 \times 10^{10}\,{\rm M}_{\odot}$ is associated with the widespread gas reservoir, most likely consisting of tidal debris, that surrounds the Dragonfly Galaxy. An additional M$_{\rm H2} \sim 2.2 \pm 0.2 \times 10^{10}\,{\rm M}_{\odot}$ is found in the central host galaxy.

The CO(1-0) in the tidal debris appears to be spread across a total extent of roughly 60 kpc. This is larger than the extent of tidal CO features observed in other high-$z$ galaxies. For example, observations with the Plateau de Bure Interferometer showed CO(3-2) emission on scales of $\sim 8-20$ kpc in two high-$z$ SMGs \citep{tac08}, while CO(2-1) observations with the Very Large Array showed tidal debris across 5 kpc in a $z=4.4$ QSO \citep{rie08}. In fact, detecting cold gas on scales of even a few tens of kpc is still rare at high-$z$. 
While ALMA observations of the bright atomic [C\,{\sc II}]$_{\rm 158 \mu m}$ have not detected any extended cold gas \citep[e.g.][]{wan13,bre14,rie14}, \citet{cic15} recently reported [C\,{\sc II}] emission spread out to $\sim$30\,kpc from the center in a $z=6.4$ QSO. The latter detection was obtained with the most compact configurations of the Plateau de Bure Interferometer. We therefore argue that sensitive short-baseline observations are key to detecting such very extended gas reservoirs. In this respect, the relatively large number of short ($<$200m) baselines that were included in our ATCA observations likely provided the sensitivity for tracing CO(1-0) emission on tens of kpc scales. 


On the other hand, the very extended molecular gas reservoir around the Dragonfly Galaxy may also serve as an extreme example that gas-rich galaxy mergers and interactions may be a common triggering mechanism for the powerful radio sources in HzRGs. This would be analogue to what is believed to occur for only the most powerful (high-excitation) radio sources at low- and intermediate-$z$, which are thought to be fueled by cold gas deposited after a major merger event \citep[e.g.][]{har07,emo10,smo11,jan12,ram12,mao14}. Of course, our current work is based on a single object and cannot exclude that other possible mechanisms may provide fuel for starformation and AGN activity in HzRGs \citep[see e.g.][]{hat13}. Nevertheless, our CO(1-0) results on the Dragonfly Galaxy provide direct observational evidence of a gas-rich (`wet') major-merger associated with a powerful high-$z$ radio galaxy. This is in agreement with the claim by \citet{ivi12} that violent gas-rich galaxy interactions are ubiquitous among starbursting HzRGs.


\subsection{Star formation $\&$ evolution}
\label{sec:SF}

\citet{dro14} show that the Dragonfly Galaxy is the most IR-luminous HzRG known in the southern hemisphere, with $L_{\rm IR\,(8-1000\,\mu m)} \sim 2.2 \times 10^{13} L_{\odot}$, which is in the regime of HyLIRGs. About 80$\%$ of the IR emission is estimated to arise from the dust-enshrouded starburst component (i.e. $L_{\rm FIR-SB} \sim 1.8 \times 10^{13} L_{\odot}$), while the remaining 20$\%$ comes from the buried AGN. Following \citet{dro14}, this corresponds to a star formation rate of SFR\,$\sim$\,3000 M$_{\odot}$\,yr$^{-1}$. 
The SFR is a factor of a few higher than any other HzRG at $z \la 2.5$ studied by \citet{dro14}, and only rivaled by three known northern HzRGs at $z \ga 3.5$. For example, the estimated starburst-IR luminosity of the Dragonfly Galaxy is more than 3$\times$ that of the massive Spiderweb Galaxy at $z=2.1$ and comparable to that of 4C41.17 at $z=3.6$ \citep{dro14}. The Dragonfly Galaxy is also among the most massive HzRGs known at any $z$, with M$_{*}$\,=\,5.8\,$\times$\,10$^{11}$ M$_{\odot}$ \citep{bre10}. 

As shown in Fig.\,\ref{fig:spectrum} and Table \ref{tab:spec}, strong \lya\ emission is detected in the Dragonfly Galaxy. The total \lya\ luminosity covered by the 1.3$''$ wide slit is $L_{\rm Ly\alpha} = 8.2 \pm 0.8 \times 10^{43}$\,erg\,s$^{-1}$. This is within the range of $L_{\rm Ly\alpha}$ that \citet{rot97} detect among a sample of 30 HzRGs at $z>1.9$. Evidence that part of the $L_{\rm Ly\alpha}$ in the Dragonfly Galaxy originates from enhanced star formation comes from the high \lya/He\,{\small II} emission-line ratio of 25 (Table \ref{tab:spec}). According to \citet{vil07lya}, a line ratio of \lya/He\,{\small II}\,$\geq$\,15 indicates that simple AGN photo-ionisation models are no longer able to explain the \lya\ content, and instead a significant fraction of the \lya\ emission is ionised by enhanced star formation.\footnote{An enhancement in the collisionally excited \lya\ and C\,{\small IV} lines with respect to He\,{\small II} can also occur from high gas densities ($>$1000 cm$^{-3}$), but this would imply that the \lya\ emission is mostly unobscured, which is not likely for the Dragonfly Galaxy. To confirm that enhanced star formation causes the large \lya/He\,{\small II} ratio, deeper optical spectra are required to search for stellar absorption features similar to those found in 4C\,41.17 \citep[][]{dey97}.} The \lya/He\,{\small II} = 25 found in the Dragonfly Galaxy is only matched by three other HzRGs out of a sample of 54 (4C\,41.17, PKS\,B1243+036 and TN\,J0205+2242), which are all three at $z > 3.5$ \citep{vil07lya}. In addition, the corresponding $L_{\rm Ly\alpha}$/$L_{\rm IR-SB} \sim 1.2 \times 10^{-3}$ is an order of magnitude below the estimated $L_{\rm Ly\alpha}$/$L_{\rm FIR} \sim 0.03$ that \citet{hum11} derive from the expected intrinsic \lya\ luminosity of massive starbursts. Combined with the fact that \lya\ has a contribution from gas that is photoionised by the AGN, it is clear that the \lya\ emission due to stellar ionisation is much fainter than expected for a system with such intense star formation. This means that the starburst in the Dragonfly Galaxy must be very heavily obscured. These \lya\ results support that the Dragonfly Galaxy is an extreme $z \sim 2$ starbursting system, in which the enhanced star formation is presumably triggered by the merger event. 

When comparing the starburst IR luminosity to the CO luminosity, $L_{\rm FIR}$/$L^{\prime}_{\rm CO}$ $\approx$ 300 $L_{\sun}$ (K km/s pc$^{-2}$)$^{-1}$ for the Dragonfly Galaxy. This is comparable to what is observed in the IR-brightest high-$z$ sub-millimetre galaxies \citep[SMGs; e.g.][]{gre05,gen10,ivi11,bot13}. The specific star formation rate of the Dragonfly Galaxy, sSFR\,=\,SFR/M$_{*}$\,=\,5.2 Gyr$^{-1}$, is also comparable to the sSFR\,$\sim$\,$1-10$\,Gyr$^{-1}$ estimated for SMGs and high-$z$ dust-obscured galaxies (DOGs) by \citet{bus12}. These estimates are higher than those obtained by cosmological simulations of SMGs undergoing smooth gas accretion \citep[e.g.][]{dav10}, leading \citet{bus12} to conclude that the high sSFRs in these systems are driven by galaxy mergers. For the Dragonfly Galaxy, with SFR\,$\sim$\,3000 M$_{\odot}$\,yr$^{-1}$ and M$_{\rm H2} \sim 5.3 \times 10^{10} {\rm M}_{\odot}$, the estimated minimum gas-depletion time-scale is t$_{\rm depl} = {\rm M}_{\rm H2}/{\rm SFR}$ $\sim 18$ Myr (assuming that the complete reservoir of cold molecular gas is efficiently converted into new stars). If the bulk of the star formation occurs in the central host galaxy, with M$_{\rm H2} \sim 2.2 \times 10^{10}\,{\rm M}_{\odot}$ of cold molecular gas available, this would be t$_{\rm depl} \sim 7$ Myr. These estimates are at the low end of the gas-depletion time-scales typically derived for high-$z$ SMGs \citep[][but see also \citealt{swi06}]{gre05,tac08,ivi11,rie11SMGSB}. They are also about an order of magnitude shorter than the gas depletion time-scales derived for a sample of starforming galaxies (SFGs) at $z = 1-3$ \citep{gen10}, or for a similar sample of IR-selected SFGs galaxies at $z$\,$\sim$\,2, which dominate the ULIRG-mode at high-$z$ through gas-accretion processes other than major mergers \citep{dad07,rod11}. This is again consistent with the interpretation that the current HyLIRG-phase of the Dragonfly Galaxy is a short period of maximum starburst activity triggered by a major gas-rich merger. As such, the Dragonfly Galaxy resembles the well-studied gas-rich galaxy group BRI 1202-0725 at $z=4.7$, which contains a merging dust-obscured SMG and FIR-hyperluminous quasar \citep[e.g.][]{wag12,sal12,car13BRI}. However, with a separation of $\sim$26 kpc between the SMG and quasar in BRI 1202-0725, the notable difference is that the Dragonfly Galaxy appears in a much more advanced stage of the merger.

Interestingly, the minimum time-scale for the star formation to deplete the cold gas is of the same order as the expected lifetime of the radio source \citep[e.g.][]{liu92,blu00}. This short depletion time-scale implies that $-$in principle$-$ the star formation itself could be the dominant mechanism for clearing the central host galaxy of cold, self-gravitating molecular gas, without the explicit need for strong negative feedback from the quasar nucleus \citep[as often invoked for high-$z$ QSOs; e.g.][]{spr05,mat05}. This short and vigorous starburst episode may perhaps indicate that we are witnessing a crucial phase of transformation that will quickly turn the Dragonfly Galaxy into a massive high-$z$ quiescent elliptical \citep[e.g.][]{san13}, once most of the cold gas has been consumed. 

Still, despite the estimated short t$_{\rm depl}$ for the current starburst event (which is most likely concentrated towards the central host galaxy), it is likely that the star formation rates within the widespread tidal debris are substantially lower than 3000 M$_{\odot}$\,yr$^{-1}$. Given that roughly 60$\%$ of the cold molecular gas is spread on scales of this tidal debris, it is likely that part of this gas reservoir will remain intact for much longer than t$_{\rm depl}$ and may trigger future, perhaps less intense, episodes of star formation, once the cold gas is re-accreted back onto the host galaxy.

\subsection{Radio source $\&$ feedback}
\label{sec:jetsISM}

The Dragonfly Galaxy has a fairly small radio source (2.2$^{\prime\prime}$/18\,kpc in diameter; \citealt{pen00radio}). On scales of a few tens of kpc, there is an apparent alignment between part of the widespread CO(1-0) emission and the radio-jet axis (Fig.\,\ref{fig:mrc0152color}). This is most notable for the CO-emission NW of the host galaxy. Unless this is merely a chance alignment, it suggests that there is a physical link between the propagating radio source and the widespread molecular gas. Intriguing support for this scenario comes from the recent discovery of cold molecular CO(1-0) reservoirs found in the halo environment of several HzRGs, with the CO(1-0) emission aligned along the radio axis and found just outside the brightest edge of the radio source \citep{emo14}.

The widespread CO(1-0) emission in the Dragonfly Galaxy suggests that it may be common for cold gas and dust to be expelled as tidal debris during the early evolution of high-$z$ radio galaxies. 
As the jets expand into this cold gas and dust, on the one hand their radio flux may be boosted, which may potentially introduce a bias in identifying these systems as HzRGs \citep{bar96,tad11}. On the other hand, it is also likely that slow shocks will develop in the multi-phase medium, leading to compression of the atomic and molecular phases of the gas \citep{gui09}. Small amounts of dust surviving in the shocked gas may catalyze the formation of molecules. The cooling provided by (predominately) H$_2$ may lead to large quantities of cold molecular gas \citep{gui10}, as observed along the radio jets of several HzRGs \citep{kla04,emo14}. Thus, the tidal stripping of gas and dust during a galaxy merger may possibly play a key role in enabling the formation of cold molecular gas along the radio jet as it expands. 

Potentially, this scenario may even trigger jet-induced star formation, which has been invoked as a possible explanation for alignments seen between radio sources and UV/optical emission in HzRGs \citep[][see also follow-up work by e.g. \citealt{ste14,zin13,kar14}]{mcc87,cha87}. In this respect, it is interested to note that the radio source in the Dragonfly Galaxy also aligns with the main axis of the host galaxy in the {\it HST} imaging of Fig.\,\ref{fig:hst}.

\section{Conclusions}

We have provided evidence that the Dragonfly Galaxy (MRC\,0152-209) is a high-$z$ `wet' merger, observed roughly 3.5 Gyr after the Big Bang. With its extreme star formation and evolving radio source, the Dragonfly Galaxy allows us to witness a short but crucial stage in the early evolution of massive galaxies, during which the mass build-up is expected to be at a maximum for a period of a few tens of Myr. This paper highlighted the crucial role that the cold molecular gas is expected to play in this.

\vspace{1.5mm}

\noindent The main results of this study are:

\vspace{1.5mm}

\noindent {\sl i).} {\it HST} imaging shows the presence of large-scale tidal features that identify the Dragonfly Galaxy as a merger.

\vspace{1.5mm}

\noindent {\sl ii).} ATCA CO(1-0) mapping reveals that cold molecular gas is associated with the tidal debris, classifying the Dragonfly Galaxy as a `wet' merger. We conclude that:

\vspace{1mm}

\hangindent=0.3cm \hspace{-0.6cm} - the cold molecular gas is distributed across a total extent of $\sim$60 kpc, making the Dragonfly Galaxy one of very few high-$z$ objects in which the CO has been imaged on these scales;

\vspace{1mm}

\hangindent=0.3cm \hspace{-0.6cm} - roughly 60$\%$ of the CO(1-0) emission (M$_{\rm H2} \sim 3 \times 10^{10}$\,M$_{\odot}$) appears to be associated with this widespread tidal debris, while the remaining 40$\%$ (M$_{\rm H2} \sim 2 \times 10^{10}$\,M$_{\odot}$) is found in the central host galaxy;  

\vspace{1mm}

\hangindent=0.3cm \hspace{-0.6cm} - the minimum gas depletion time-scale is a short $7-18$ Myr (although the gas reservoir at large scales may trigger future star formation episodes, once re-accreted back onto the galaxy);

\vspace{1mm}

\hangindent=0.3cm \hspace{-0.6cm} - the Dragonfly Galaxy may be a high-$z$ analogue to the most powerful radio sources at low-$z$, whose activity is believed to be fueled by cold gas deposited after a galaxy merger.

\vspace{1.5mm}

\noindent {\sl iii).} Consistent with the very high star formation rates (SFR\,$\sim$\,3000 M$_{\odot}$ yr$^{-1}$) found in previous IR studies of the Dragonfly galaxy, our new WHT spectrum confirm the presence of enhanced and heavily obscured star formation through a high \lya/He\,{\small II} emission-line ratio and low $L_{\rm Ly\alpha}/L_{\rm IR-SB}$.

\vspace{1.5mm}

\noindent {\sl \noindent iv).} The small (18 kpc) radio source appears to be aligned with widespread CO(1-0) emission, suggesting a physical link between the propagating radio jets and the cold gas reservoir. We propose a scenario where tidally ejected gas and dust that is overrun by a radio jet may undergo compression by slow shocks, resulting in the formation and cooling of molecular gas along the radio axis.



\vspace{1.5mm}

\noindent Concluding, as the most IR-luminous southern HzRG, the Dragonfly Galaxy provides us with a unique opportunity to further investigate the most active stages of galaxy evolution in the Early Universe.

\section*{Acknowledgments}
We thank Santiago Arribas for useful discussions. We also thank the anonymous referee for valuable suggestions that significantly improved this paper. BE gratefully acknowledges that the research leading to these results was funded by the European Union Seventh Framework Programme (FP7-PEOPLE-2013-IEF) under grant agreement n$^{\circ}$\,624351, and by Australia's Commonwealth Scientific and Industrial Research Organisation (CSIRO). MVM's work has been funded with support from the Spanish Ministerio de Econom\'{i}a y Competitividad through the grant AYA2012-32295. NS is the recipient of an ARC Future Fellowship. We thank the staff at the Australia Telescope Compact Array for their dedicated help in making our ATCA project successful. The Australia Telescope is funded by the Commonwealth of Australia for operation as a National Facility managed by CSIRO. Based on observations made with the NASA/ESA Hubble Space Telescope, obtained at the Space Telescope Science Institute, which is operated by the Association of Universities for Research in Astronomy, Inc., under NASA contract NAS 5-26555 -- these observations are associated with program $\#$8183. The Wiliam Herschel Telescope is operated on the island of La Palma by the Isaac Newton Group in the Spanish Observatorio del Roque de los Muchachos of the Instituto de Astrofísica de Canarias. The National Radio Astronomy Observatory is a facility of the National Science Foundation operated under cooperative agreement by Associated Universities, Inc.


\end{document}